\documentclass[preprint,sort&compress,12pt]{elsarticle}

\usepackage{graphicx}
\usepackage{amsmath}
\usepackage{bm}
\usepackage{lineno}
\usepackage{subfigure}
\usepackage{url}
\usepackage[colorlinks,unicode=true]{hyperref}
\usepackage{rotating}
\usepackage{epstopdf}
\usepackage{float}

\journal{Nuclear Inst. and Methods in Physics Research, A}

\makeatletter
\def\ps@pprintTitle{
\def\@oddhead{\copyright\,2018. This manuscript is available under license CC BY-NC-ND 4.0 \hfill}}
\makeatother

\begin{document}

\begin{frontmatter}
\title{Timing analysis of Swift J1658.2--4242's outburst in 2018 with \emph{Insight}-HXMT, \emph{NICER} and \emph{AstroSat}}

\author[a1,a2]{Guangcheng Xiao}
\ead{xiaogc@ihep.ac.cn}
\author[a1]{Yu Lu}
\author[a1]{Xiang Ma}
\author[a1]{Mingyu Ge\corref{cor1}}
\ead{gemy@ihep.ac.cn}
\author[a3]{Linli Yan}
\author[a4]{Zijian Li}
\author[a1,a2]{Youli Tuo}
\author[a1,a2]{Yue Zhang}
\author[a1,a2]{Wei Zhang}
\author[a1,a2]{Hexin Liu}
\author[a1,a2]{Dengke Zhou}
\author[a1]{Liang Zhang}
\author[a1]{Qingcui Bu}
\author[a1]{Xuelei Cao}
\author[a1]{Weichun Jiang}
\author[a1]{Yupeng Chen}
\author[a1,a2]{Shu Zhang}
\author[a5]{Li Chen}
\author[a1,a2]{Jinlu Qu}
\author[a1,a2]{Liming Song}
\author[a1,a2]{Shuangnan Zhang}

\address[a1]{Key Laboratory of Particle Astrophysics, Institute of High Energy Physics, Chinese Academy of Sciences, Beijing 100049, China}
\address[a2]{University of Chinese Academy of Sciences, Beijing 100049, China}
\address[a3]{School of Mathematics and Physics, Anhui Jianzhu University, Hefei 230601, China}
\address[a4]{School of Architecture and Art, Hebei University of Engineering, Handan 056038, China}
\address[a5]{Department of Astronomy, Beijing Normal University, Beijing 100875, China}
\cortext[cor1]{Corresponding author.}

\begin{abstract}
We present the observational results from a detailed timing analysis of the black hole candidate Swift J1658.2--4242 during its  2018 outburst with the observations of {\it Hard X-ray Modulation Telescope} ({\it Insight}-HXMT), {\it Neutron Star Interior Composition Explorer} ({\it NICER}) and {\it AstroSat} in 0.1--250\,keV. The evolution of intensity, hardness and integrated fractional root mean square (rms) observed by {\it Insight}-HXMT and {\it NICER} are presented in this paper. Type-C quasi-periodic oscillations (QPOs) observed by {\it NICER} (0.8--3.5\,Hz) and {\it Insight}-HXMT (1--1.6\,Hz) are also reported in this work. The features of the QPOs are analysed with an energy range of 0.5--50\,keV. The relations between QPO frequency and other characteristics such as intensity, hardness and QPO rms are carefully studied. The timing and spectral properties indicate that Swift J1658.2--4242 is a black hole binary system. Besides, the rms spectra of the source calculated from the simultaneous observation of {\it Insight}-HXMT, {\it NICER} and {\it AstroSat} support the Lense-Thirring origin of the QPOs.  The relation between QPO phase lag and the centroid frequency of Swift J1658.2--4242 reveals a near zero constant when $< 4$\,Hz and a soft phase lag at 6.68\,Hz. This independence follows the same trend as the  high inclination galactic black hole binaries such as MAXI J1659--152.
\end{abstract}

\end{frontmatter}



\section{Introduction}           
\label{sect:intro}

During a series of X-ray missions since the last half centuries, a set of X-ray transients were found among which about 20 sources are confirmed as black holes and 30 as black hole candidates. The black hole transients (BHTs) usually go through similar state transitions during an outburst \citep[][]{Remillard2006, van2006}. By analysing the timing and spectral features of different stages of an outburst,  states such as low-hard state (LHS), hard-intermediate state (HIMS), soft-intermediate state (SIMS) and high-soft state (HSS) are distinguished \citep{Belloni2016}. In the LHS, the emission is dominated by a strong comptonized emission and the intrinsic variability can reach 30--40\% \citep{Mendez1997}. The HIMS and SIMS usually have both significant hard component and the contribution of a thermal disk. The variability can vary a lot and normally ranges between 5--20\%. The HSS is dominated by a strong thermal component associated with an accretion disk which may extend to the innermost stable circular orbit (ISCO). The variability is very low which is often consistent with zero \citep{Motta2016}. Three main types of low frequency quasi-periodic oscillations (LFQPOs) namely Type-A/B/C QPOs are classified according to their intrinsic properties (mainly centroid frequency, variability amplitude and width), the underlying broad-band noise components (noise shape and total variability level) and the relations among these quantities \citep{Motta2016}. The existing models that attempt to explain the origin of LFQPOs are generally based on two different mechanisms: instabilities and geometrical effects \citep{Motta2015}. In the former case, the transitions of instabilities are often concerned in unique accretion geometry \citep[see e.g.][]{Tagger1999, Titarchuk2004, Cabanac2010}. In the latter case, the physical process is typically related to precession \citep{Ingram2009}. Recent years, thanks to the accumulation of X-ray data (especially {\it Rossi X-ray Timing Explorer--RXTE}), a global statistical and analytical timing researches have achieved satisfactory progress. \citet{Motta2015} analysed the relation between the rms of LFQPOs and the orbital inclination in 14 black hole X-ray transients, getting the evidence of geometrical origin of LFQPOs. \citet{Eijnden2017} studied the relation between the phase lag at LFQPOs (harmonics and subharmonics as well if exist) and the orbital inclination of black hole transients which again indicated the Lense-Thirring precession model accounting for Type-C QPO. In addition, \citet{Gao2017} studied the timing and spectral properties of 99 Type-B QPOs from eight black hole X-ray binaries and classified two groups which may differ in the geometry of corona. While the origins of Type-A/B QPOs  still remain uncertain so far.

The X-ray transient source Swift J1658.2--4242 was first detected by {\it Swift}/BAT on February 16, 2018 (see GCN \#22416, GCN\#22417). The following-up pointing observation by {\it Swift}/XRT improved the position and revealed the possible optical counterpart \citep{Mereminskiy2018}. Also the source was detected by {\it INTEGRAL}(IBIS/ISGRI) during its observations of the Galactic center field on 2018 February 13 \citep[][]{Grebenev2018}. Besides, the above {\it Swift}/XRT observation revealed that the spectrum of the source can be described by a strongly absorbed power-law with a photon index of $1.2\pm0.3$ and an absorption column density of $\rm{N_H} = 1.5 \times 10^{23}\,\rm{cm^{-2}}$. Meanwhile, the {\it NuSTAR} spectrum of the source in the hard state can be fitted with an absorbed cutoff powerlaw model. The spectral fit of {\it NuSTAR} gave a similar absorption column density of $\rm{N_H} = 1.3 \times 10^{23}\,\rm{cm^{-2}}$, a photon index of 1.3 and a cut-off energy of 53\,keV. LFQPOs whose central frequency evolving from $0.138 \pm 0.002$\,Hz to $0.166 \pm 0.003$\,Hz was reported in the {\it NuSTAR} observation\citep{Xu2018Atel}. Later, four days after the {\it NuSTAR} observation, the {\it AstroSat}/LAXPC spectral fit gave a consistent result of an absorption column density of $\rm{N_H} = (1.6 \pm 0.2) \times 10^{23}\,\rm{cm^{-2}}$, a power law photon index of $1.76 \pm 0.06$ , a high-energy cutoff of $46 \pm 3$\,keV and an inner disk temperature of $1.4 \pm 0.1$\,keV. Besides, the {\it AstroSat}'s Large Area X-ray Proportional counters (LAXPC) power density spectral (PDS) showed the presence of band-limited noise coupling strong sharp QPO (rms of $\sim$16\%), whose centroid frequency increased with time from $\sim$1.6 to 2\,Hz. The total 0.1--20\,Hz fractional rms is $\sim$34\% over 3--80\,keV. These features indicate that the source was in the hard intermediate state \citep{Beri2018}. All observations performed by the X-ray missions mentioned above propose Swift J1658.2-4242 to be a black hole candidate.

At the same time, {\it Insight}-HXMT carried out a series of pointing observations which suppled us a set of observations with high time resolution and broad energy band (1--250\,keV). The QPOs whose central frequency ranging form 1\,Hz to 1.6\,Hz were observed in this outburst by {\it Insight}-HXMT. We also bring in the {\it NICER} observations for timing analysis because {\it NICER} has better performance in lower energy band (0.1--12\,keV) and its observations on Swift J1658.2-4242 spread for a longer time interval. Meanwhile, we also go further into a nearly simultaneous QPO observation of {\it AstroSat}/LAXPC, computing the QPO rms spectrum in order to make a comparison with {\it Insight}-HXMT.

In this work, we present a detailed timing analysis of Swift J1658.2--4242 in its 2018 outburst with the observations of {\it Insight}-HXMT, {\it NICER} and {\it AstroSat}. The structure of this work is described as follows. Section \ref{sect:ObservationsAndDataReduction} gives an introduction of the {\it NICER}, {\it Insight}-HXMT and {\it AstroSat} observations and summarizes the data reductions of the three main telescopes deployed on {\it Insight}-HXMT. The timing results are presented in section \ref{sect:Results}. An overview of the whole outburst is given at first. Then the evolution of QPO frequency, the relations between QPO frequency and other timing quantities are studied. A simultaneous observation in which QPOs are detected by {\it Insight}-HXMT, {\it NICER} and {\it AstroSat} is presented. At the end of the section, the humps observed by {\it Insight}-HXMT are listed in this section. In Section \ref{sect:DiscussionAndSummary}, the physical implications of our results are discussed. Through the paper, uncertainties are given at the 68\% confidence level.

\section{Observations and Data reduction}
\label{sect:ObservationsAndDataReduction}

\subsection{Observations}
\label{subsect:Observations}
{\it Insight}-HXMT, China's first X-ray astronomical satellite which was successfully launched on June 15, 2017, carries three sets of main instruments (LE/ME/HE, short for the Low/Medium/High Energy X-ray Telescope respectively). The detectors of LE are the Swept Charge Devices (SCDs). ME is composed of 1728 Si-PIN detectors (1728 pixels for simplicity). HE contains 18 cylindrical NaI(Tl)/CsI(Na) phoswich detectors. The major capabilities of these three main detectors are summarised in Table~\ref{tab:HXMTInfo}. The detailed information can be found in \citet{Zhang2018}.
\begin{table}[h]
\begin{center}
 \caption{The Capabilities of 3 Main Detectors on Board {\it Insight}-HXMT.}
 \label{tab:HXMTInfo}
 \begin{tabular}{lccc}
  \hline
  \hline
  Detectors          &      LE        &      ME      &    HE            \\
  \hline
  Type               &      SCD       &    Si-PIN    &    NaI/CsI       \\
  Area($\rm{cm}^2$)  &      384       &     952      &    5000          \\
  Energy Range       &   1--15\,keV     &  5--30\,keV    &    20--250\,keV    \\
  Time Resolution    &     1\,ms       & 280\,$\mu$s   &   25 $\mu$s      \\
  Energy Resolution  & 2.5\% @\,6\,keV   & 14\% @\,20\,keV &   19\% @\,60\,keV   \\
  \hline
  \hline
 \end{tabular}
 \end{center}
\end{table}

From February 20 to March 21 in 2018, {\it Insight}-HXMT performed 47 pointing observations on Swift J1658.2--4242 with a net exposure of $\sim$500\,ks. During February 17 to August 21 in 2018, {\it NICER} carried out 42 pointing observations on Swift J1658.2--4242. These observations provide a view of the whole outburst. Though not continuously, {\it AstroSat} carried out 25 orbit-cycle observations to Swift J1658.2--4242. We searched all the observations of {\it AstroSat} for QPOs and present them in this work.

As listed in Table~\ref{tab:HXMTInfo}, the total effective area of LE detectors is much smaller than that of ME/HE. Besides, the stricter criteria for filtering data (see subsection~\ref{subsect:DataReduction}) makes the net exposure of LE much less than that of ME/HE. In addition, the contamination of other bright sources in the field of view (FoV) of the instrments makes the situation even worse. When dealing with this relatively faint source Swift J1658.2--4242, we find that the unsatisfactory signal-to-noise ratio of LE makes it hard to detect the QPO signal. Compared with LE, the X-ray Timing Instrument (XTI) of {\it NICER} provides an effective area of 1900\,cm$^{2}$ at 1.5\,keV \citep{Gendreau2016, Prigozhin2012}. The {\it NICER} observations are suitable for analysing the properties of QPOs in 1--10\,keV energy band. As for {\it AstroSat}, LAXPC provides very large 6100/4500/5100\, $\rm{cm}^2$ effective areas at 10/30/50\,keV \citep[and references therein]{Yadav2016}, which results in higher significance of the QPO signal detected by LAXPC compared with XTI and ME in the same energy bands. Through out the paper, the features in 1--10 keV of Swift J1658.2--4242 are mainly obtained from {\it NICER} and {AstroSat} considering the above situations.

\subsection{Data reduction}
\label{subsect:DataReduction}
Data reduction of the {\it Insight}-HXMT observations is done by the {\it Insight}-HXMT Data Analysis software (HXMTDAS) v2.0. The processes are summarized into five steps for the three main instruments: (1) converting the photon channel to PI; (2) generating good time intervals (GTIs) for each detectors with stand criteria; (3) reconstructing the split events for LE and calculating the grade and dead time of ME events; (4) selecting the events with GTIs; (5) using the screened events to generate lightcurves and backgrounds.

The criteria for filtering the data of the three main instruments are: (1) pointing offset angle $<0.05^\circ$; (2) elevation angle $>6^\circ$; (3) value of the geomagnetic cutoff rigidity $>6$ \citep{HuangYue2018}. Since LE detectors are much more sensitive to bright earth, we set the bright earth elevation angle (DYE\_ELV) $>40^{\circ}$ which is higher than that of ME/HE (DYE\_ELV $>0^{\circ}$). This data selection makes the net exposure time of LE detectors less than ME/HE. The background lightcurves are generated by the latest background software. The count rate of blind FoV detectors are used to extract the background by a rough proportional correction $B = N*C_b$ when generating the net lightcurves, where $B$ stands for background count rate of the selected data, $C_b$ stands for the rate of blind FoV detectors and $N$ for the proportional coefficient \citep{HuangYue2018}. We chose only the small FoV detectors for LE/ME because the background of large FoV detectors is much higher.

The lightcurves of {\it NICER}/XTI are generated with cleaned  events data by software \textbf{XSELECT}. The lightcurves, spectra and background of {\it AstroSat}/LAXPC are generated by the LAXPC individual routines software (May 19th 2018 version) provided by Indian Space
Research Organisation (ISRO). The data of LAXPC10 and 20 cameras are used in this work, with LAXPC30 excluded due to a suspected gas leak leading to a loss of efficiency~\citep{Beri2018}.

\subsection{Analysis methods}
\label{subsect:methods}
As a generally  used method for black hole candidates (BHCs) to characterize the spectral properties of a source, the hardness-intensity diagram (HID) is often adopted to visually present the accretion states. Many black hole transients evolve  with a `q' shape in the HID when going through a whole `canonical outburst' \citep[see e.g.][]{Altamirano2015, Homan2001}. In this work, the hardness is defined as the count rate ratio between 6--10\,keV and 1--6\,keV both for LE and {\it NICER}. Besides, the hardness-rms diagram (HRD) is adopted in this paper because it can provide a reference for each state.

We extract lightcurves with $2^{-5}$\,s time resolution from the observations done by XTI, ME and HE in energy bands 1--10\,keV, 6--30\,keV and 25--50\,keV to compute the power density spectra using Leahy normalization \citep{Leahy1983} with \textbf{powspec} in \textbf{XRONOS}. The PDSs are fitted with a model consisting of a constant Poisson noise and several Lorentzians accounting for the possible broad band noise (BBN), the possible QPO fundamental and (sub)harmonics. The fractional rms of QPO for Leahy normalization is given by
\begin{equation}
    rms_{\rm QPO}=\sqrt{\frac{\pi \times LW \times LN }{2 \times (S + B)}} \times \frac{S + B}{S},
	\label{eq:rmsQPO}
\end{equation}
where $LW$ is the full width at half-maximum (FWHM) of the QPO, $LN$ is the normalization of Lorentzian component for QPO, $S$ is the source count rate and $B$ is the background count rate~\citep{Bu2015}. The background correction in equation (\ref{eq:rmsQPO}) is necessary for {\it Insight}-HXMT in the case of Swift J1658.2--4242 because its background level in ME is comparable with the source level and even 2 times higher than the that of the source in HE. The same correction is applied for the LAXPC, for the total background photon number is about 1/3 of the source in the case of Swift J1658.2--4242. When dealing with {\it NICER} observations, the correction can be ignored because even taking the total count rate in quiescent state as background of QPO observations, the correction is less than 4\%. After subtracting the Poisson noise, we integrate the rms within 0.1--16\,Hz as total rms. The significance of QPOs is given as the ratio of the integral of the power of the Lorentzian used to fit the QPO divided by the negative 1\,$\sigma$ error on the integral of the power \citep{Motta2015}.

To obtain the phase lag at QPO frequency, the complex cross-power spectrum  are calculated between the soft and hard energy band lightcurves. Then the real and imaginary part are averaged separately within the QPO frequency range ($f_{\rm QPO}\pm{\rm FWHM/2}$). The details for calculating cross-power spectrum and phase lag can be found in the work of \citet{Nowak1999}.

\section{Results}
\label{sect:Results}
\subsection{Overview of the outburst}
\label{subsect:Overview}

\begin{figure}[h]
  \centering
  \includegraphics[width=0.7\columnwidth]{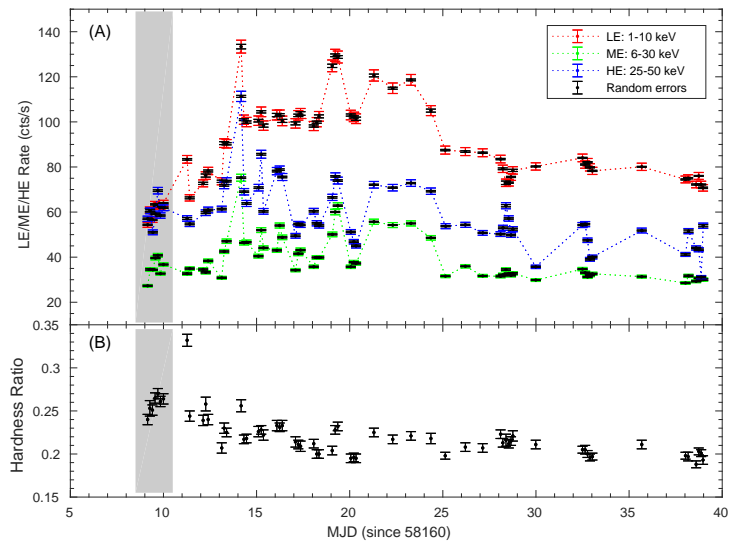}
  \caption{X-ray lightcurves and hardness evolution diagram of Swift J1658.2--4242 in its 2018 outburst observed by {\it Insight}-HXMT. Panel (A) presents the lightcurves by LE lightcurve (1--10\,keV), ME lightcurve (6--30\,keV) and HE lightcurve (25--50\,keV). A systematical error of 0.02 is add to the lightcurves of three instruments. The random errors are presented in the lightcurves with black-dot symbol. The hardness in panel (B) is defined as the count rate ratio between 6--10\,keV and 1--6\,keV). The grey area corresponds to the time interval when QPOs are detected. in ME/HE lightcurves correspond to QPO observations.}
  \label{Fig:HXMTLightCurve}
\end{figure}

 The net lightcurves and hardness evolution of Swift J1658.2--4242 observed by the three main instruments on board {\it Insight}-HXMT are presented in Fig.~\ref{Fig:HXMTLightCurve}. The energy bands of LE/ME/HE are 1--10\,keV, 6--30\,keV and 25--50\,keV, respectively. A systematical error of 2\% of the background level has been added to the lightcurves of three instruments considering the uncertainties of the current calibration\footnote{The white book of {\it Insight}-HXMT suggests a systematical error of 1.7\%/2.6\%/1.3\% separately for LE/ME/HE (\underline{http://hxmt.org/index.php/2013-03-22-08-08-48/docs}).}. Lightcurves observed by these three instruments show obvious rising and falling stages of the outburst. The hardness ratio (6--10\,keV VS. 1--6\,keV) calculated from LE observations shows significant decrease at the beginning of the outburst. QPOs detected by ME/HE during this stage can be seen in the grey area of panel (A) of Fig.~\ref{Fig:HXMTLightCurve}.

\begin{figure}[h]
  \centering
  \includegraphics[width=0.7\columnwidth]{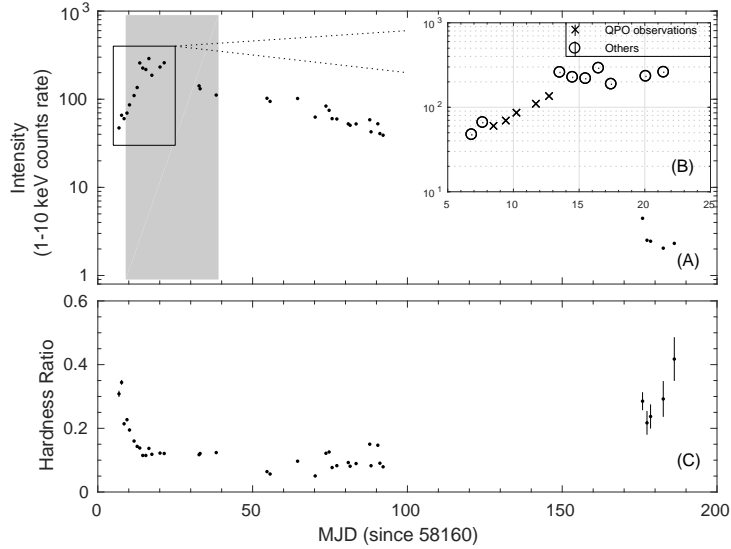}
  \caption{The lightcurve and hardness evolution diagrams of Swift J1658.2--4242's outburst in 2018 observed by {\it NICER}. Each point correspond to a {\it NICER} observation. The upper panel shows the 1--10\,keV energy band lightcurve. The grey area correspond to the interval when {\it Insight}-HXMT performed observations. There was an interval lasting about 80 days that {\it NICER} did not perform any observation before the end of the outburst. Panel (B) presents the location of QPO observations in the raising phase of the outburst. The last five observations show that the source count rate stay in the $1.5\sim2$ cts/s level at quiescent state, which can be used to evaluate the background level. The lower panel shows the hardness evolution of the outburst. The hardness is defined as the count rate ratio between 6--10\,keV and 1--6\,keV.}
  \label{Fig:NICERLightCurve}
\end{figure}

 Fig.~\ref{Fig:NICERLightCurve} presents the lightcurve and hardness evolution of Swift J1658.2--4242's outburst observed by {\it NICER}. The source experienced fast rising and slow falling stages shown in panel (A) of Fig.~\ref{Fig:NICERLightCurve}. The {\it Insight}-HXMT observations were carried out during the time represented by the grey area. Panel (B) of Fig.~\ref{Fig:NICERLightCurve} shows the observations in which QPOs are detected during the raising stage. Little information of how the source got back to LHS or quiescent state can be provided since the lack of about lasting 80-day observations. Panel (C) of Fig.~\ref{Fig:NICERLightCurve} shows the hardness evolution observed by {\it NICER}. The sharp increase in intensity and concurring sharp decrease in hardness characterize the LHS and HIMS of the source in this outburst.

\begin{figure}[h]
  \centering
   \includegraphics[width=0.7\columnwidth]{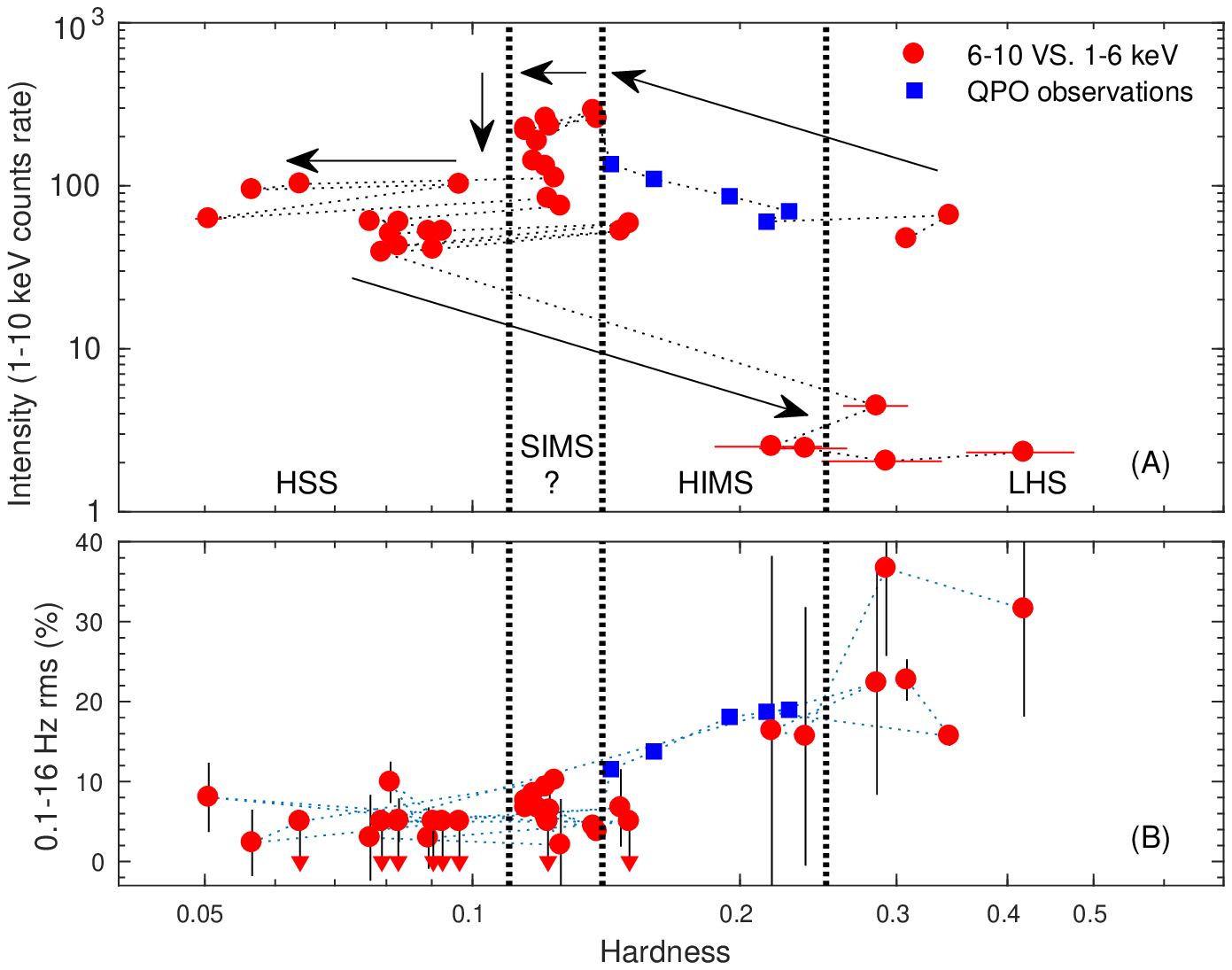}
   \caption{Panels (A) and (B) present the HID and HRD of Swift J1658.2--4242 in its 2018 outburst. Hardness is defined as the count rate ratio between 6--10\,keV and 1--6\,keV energy band of {\it NICER}. The total rms are integrated within 0.1--16\,Hz. Each point corresponds to an observation of {\it NICER}. The blue squares represent the QPO observations. The black arrows indicate the direction of time evolution.}
   \label{Fig:HIDHRDNicer}
\end{figure}

The HID and HRD of the outburst are presented in panel (A) and panel (B) of Fig.~\ref{Fig:HIDHRDNicer}, respectively. As is shown in Fig.~\ref{Fig:HIDHRDNicer}, four states are classified, mainly by the position of QPOs, the total variability amplitude, the relative hardness and intensity in this outburst. In panel (A) of Fig.~\ref{Fig:HIDHRDNicer}, the observations of {\it NICER} started when Swift J1658.2--4242 was in LHS. Then Swift J1658.2--4242 turned into HIMS very soon where low frequency Type-C QPOs were observed (see subsection~\ref{subsect:QPOobservations}). The integrated rms dropped from 20\% to 10\% when hardness dropped from 0.25 to 0.15. Then the Type-C QPO disappeared and integrated rms varied within 5--10\%, corresponding to the features often observed in SIMS. Since these is no detailed spectral analysis, the existence of SIMS remains doubtful. Then when the source turned to HSS (the hardness $< 0.1$), the integrated rms dropped to $\sim$ 5\%, which means that the source turned to soft state. The HID presents a `q' shape similar to many black-hole transients, such as GX 339--4 \citep{Motta2011}. The HRD in Fig.~\ref{Fig:HIDHRDNicer} shows that the integrated rms decreased when Swift J1658.2--4242 turned to high soft state. The features presented in the HID and HRD support the proposal that Swift J1658.2--4242 is a black hole transient candidate \citep{Xu2018B}.

\subsection{Low frequency QPOs}
\label{subsect:QPOobservations}
\begin{figure}[!ht]
  \centering
   \includegraphics[width=0.85\columnwidth]{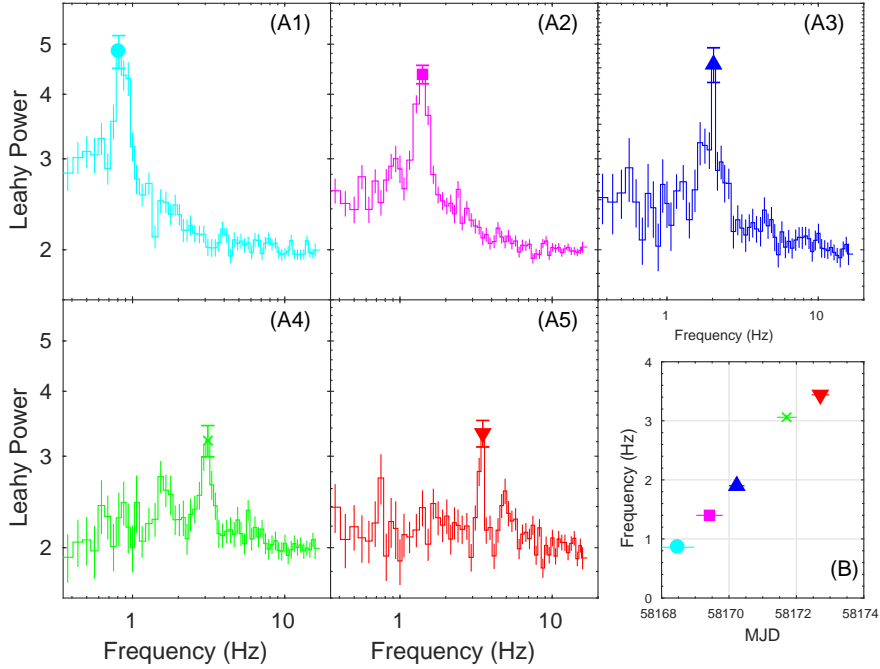}
   \caption{Panel (A1-A5) presents the QPO PDSs computed from {\it NICER} observations (1--10\,keV) whose IDs are from 1200070103 to 1200070107. Panel (B) shows the evolution of the QPO frequency.}
   \label{Fig:QPOPDSEvol}
\end{figure}

The QPOs observed by {\it NICER}, {\it AstroSat}, {\it Insight}-HXMT are summarized in Tables~\ref{tab:QPOInfoNicer} and \ref{tab:QPOInfoHXMT}, including QPO frequency, FWHM, rms and significance. The QPO rms observed by XTI, LAXPC, ME and HE are calculated in energy bands 1--10\,keV, 3--50\,keV, 10--25\,keV and 25--50\,keV, respectively. As can be seen in the two tables, Type-C QPOs are detected by three satellites on the same day (MJD 58169). Fig.~\ref{Fig:QPOPDSEvol} presents the PDSs of the {\it NICER} observations in which QPOs are detected and the evolution of QPO frequency. The chosen energy band is 1--10\,keV. As is shown in panel (B) of Fig.~\ref{Fig:QPOPDSEvol}, the QPO frequency gradually increased from 0.85\,Hz to 3.47\,Hz in four days. Usually, a signal is considered to be a QPO only when its significance is greater than $3\,\sigma$. However, the significance of the signal detected in the observations (ID: 1200070106, 1200070107) of {\it NICER} is smaller than $3\,\sigma$. However, these two observations are treated as QPOs because they correspond to the QPOs' evolution.

\begin{table}[!ht]
\begin{center}
 \caption{The QPO Information Observed by {\it NICER} and {\it AstroSat}. The chosen energy bands for {\it NICER} and {\it AstroSat} are 1--10\,keV and 3--50\,keV, respectively.}
 \label{tab:QPOInfoNicer}
 \resizebox{\textwidth}{!}{
 \begin{tabular}{lccccccc}
  \hline
  \hline
  NICER.ObsID     &    MJD    & exposure(s) & Frequency (Hz)  &    FWHM (Hz)    &   QPO rms (\%) &  Significance   \\
  \hline
  1200070103      & 58168.48  &    3587     & 0.85 $\pm$ 0.01  & 0.19 $\pm$ 0.04 & 10.5 $\pm$ 1.2 &  3.1\,$\sigma$   \\
  1200070104      & 58169.42  &    5310     & 1.40 $\pm$ 0.01  & 0.28 $\pm$ 0.03 & 11.4 $\pm$ 0.8 &  4.9\,$\sigma$   \\
  1200070105      & 58170.23  &    1384     & 1.97 $\pm$ 0.04  & 0.92 $\pm$ 0.16 & 15.2 $\pm$ 1.6 &  3.3\,$\sigma$   \\
  1200070106      & 58171.71  &    1384     & 3.06 $\pm$ 0.04  & 0.55 $\pm$ 0.12 & 9.53 $\pm$ 1.4 &  2.5\,$\sigma$   \\
  1200070107      & 58172.71  &    1449     & 3.47 $\pm$ 0.02  & 0.17 $\pm$ 0.07 & 6.10 $\pm$ 2.0 &  1.1\,$\sigma$   \\
  \hline
  AstroSat.Orbit  &           &             &                  &                 &                &                 \\
  12977           & 58169.87  &    8973     & 1.57 $\pm$ 0.004 & 0.23 $\pm$ 0.01 & 16.0 $\pm$ 0.3 &  15.7\,$\sigma$  \\
  12978           & 58169.98  &    4808     & 1.59 $\pm$ 0.004 & 0.24 $\pm$ 0.01 & 16.5 $\pm$ 0.6 &  10.5\,$\sigma$  \\
  12980           & 58170.05  &    4334     & 1.68 $\pm$ 0.004 & 0.23 $\pm$ 0.01 & 15.8 $\pm$ 0.6 &   9.3\,$\sigma$  \\
  12981           & 58170.04  &    4215     & 1.78 $\pm$ 0.004 & 0.27 $\pm$ 0.01 & 16.0 $\pm$ 0.6 &  10.0\,$\sigma$  \\
  13141 + 13142   & 58180.54  &    9161     & 6.68 $\pm$ 0.02  & 1.56 $\pm$ 0.07 &  7.2 $\pm$ 0.2 &  13.6\,$\sigma$  \\
  \hline
 \end{tabular}
 }
 \end{center}
\end{table}

\begin{table}[!ht]
\begin{center}
 \caption{The QPO Information Observed by {\it Insight}-HXMT. The chosen energy bands for ME and HE are 10--25\,keV and 25--50\,keV, respectively.}
 \label{tab:QPOInfoHXMT}
 \resizebox{\textwidth}{!}{
 \begin{tabular}{lccccccc}
  \hline
  \hline
  HXMT          &   Central &    HE       &        HE        &        HE        &      ME        &       HE        &    ME/HE       \\
  ObsID         &    MJD    & exposure (s)&  Frequency (Hz)  &     FWHM (Hz)    &  QPO rms (\%)  &   QPO rms (\%)  & Significance   \\
  \hline
  P011465800101 & 58169.09  &    2692     & 1.04 $\pm$ 0.01 & 0.18 $\pm$ 0.03 & 15.6 $\pm$ 3.3 & 18.1 $\pm$ 1.9  & 1.7 / 3.4\,$\sigma$   \\
  P011465800102 & 58169.25  &    1610     & 1.15 $\pm$ 0.02 & 0.22 $\pm$ 0.05 & 16.1 $\pm$ 7.3 & 19.6 $\pm$ 2.6  & 0.8 / 2.7\,$\sigma$   \\
  P011465800103 & 58169.39  &    3108     & 1.29 $\pm$ 0.01 & 0.26 $\pm$ 0.04 & 15.1 $\pm$ 3.0 & 21.4 $\pm$ 1.9  & 1.8 / 4.0\,$\sigma$   \\
  P011465800104 & 58169.53  &    2781     & 1.36 $\pm$ 0.01 & 0.24 $\pm$ 0.03 & 16.7 $\pm$ 3.0 & 21.1 $\pm$ 1.8  & 2.0 / 4.1\,$\sigma$   \\
  P011465800105 & 58169.66  &    2146     & 1.46 $\pm$ 0.02 & 0.19 $\pm$ 0.06 & 12.6 $\pm$ 3.8 & 14.8 $\pm$ 3.4  & 1.2 / 1.5\,$\sigma$   \\
  P011465800106 & 58169.79  &    3218     & 1.53 $\pm$ 0.01 & 0.23 $\pm$ 0.04 & 15.9 $\pm$ 3.6 & 18.9 $\pm$ 2.1  & 1.6 / 3.3\,$\sigma$   \\
  P011465800107 & 58169.95  &    3101     & 1.61 $\pm$ 0.01 & 0.29 $\pm$ 0.04 & 14.4 $\pm$ 2.5 & 21.8 $\pm$ 2.2  & 2.1 / 3.6\,$\sigma$   \\
  \hline
  \hline
 \end{tabular}
 }
 \end{center}
\end{table}

As listed in Table~\ref{tab:QPOInfoHXMT}, these seven observations by {\it Insight}-HXMT were performed on MJD 58169, corresponding to {\it NICER} observation 1200070104. Compared with that of the {\it NICER}, the observations by {\it Insight}-HXMT took a whole day. During these observations, the QPO frequency increased from 1\,Hz to 1.6\,Hz. The significance of QPOs detected by HE is relatively high, while the simultaneous ME detections show lower significance because of its smaller effective area.

\subsection{Simultaneous observation}
\label{subsect:SimuObservation}
\begin{figure}[!htb]
  \centering
   \includegraphics[width=0.7\columnwidth]{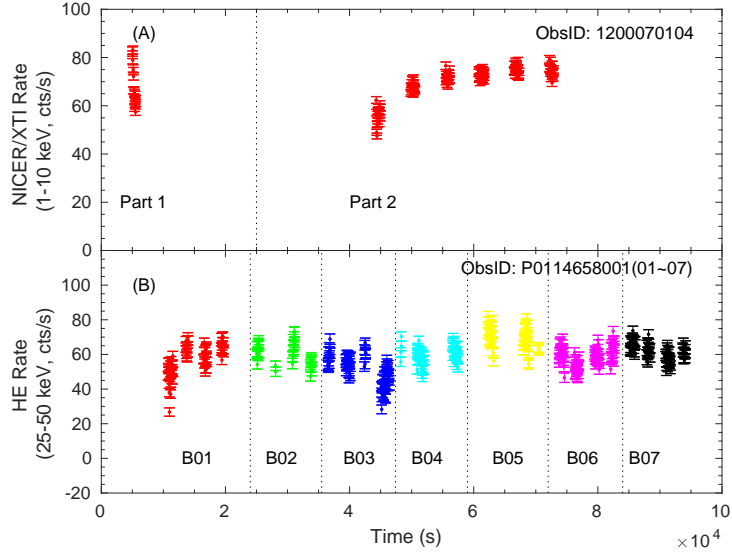}
   \caption{The simultaneous observation net lightcurves (32\,s bin) of Swift J1658.2--4242 observed by XTI and HE. The XTI lightcurve is generated from observation 1200070104 and can be split into 2 parts. The HE lightcurve (25--50\,keV) is generated from observation P0114658001 which contains 7 exposure parts (marked as `Bxx').}
   \label{Fig:SimuLightCurve}
\end{figure}

The lightcurves of the simultaneous observation generated from {\it NICER} (1200070104) and {\it Insight}-HXMT (P0114658001) are presented in Fig.~\ref{Fig:SimuLightCurve} with a 32 s time resolution. As is shown in Fig.~\ref{Fig:SimuLightCurve}, the {\it NICER} lightcurve can be split into two parts. Part 2 is observed at the same time with {\it Insight}-HXMT.

\begin{figure}[!htb]
  \centering
   \includegraphics[width=0.95\columnwidth]{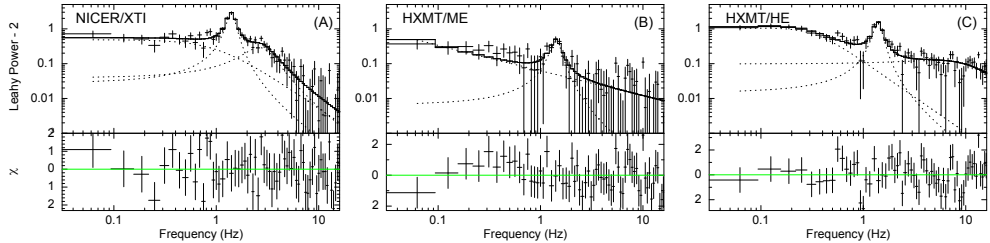}
   \caption{The QPO PDSs of Swift J1658.2--4242 calculated from the simultaneous XTI and ME/HE observations. Panels (A, B, C) correspond to XTI 3--10\,keV, ME 10--25\,keV and HE 25--50\,keV respectively. The Poisson noise is subtracted from the PDSs.}
   \label{Fig:SimuPDS}
\end{figure}

The PDSs computed from simultaneous XTI/ME/HE observations are presented in Fig.~\ref{Fig:SimuPDS}. The fit of the PDSs computed by the data of XTI (3--10\,keV), ME (10--25\,keV) and HE (25--50\,keV) gives the QPO frequency of $1.40 \pm 0.01$\,Hz, $1.42 \pm 0.03$\,Hz and $1.39 \pm 0.01$\,Hz, respectively. The FWHMs are $0.28 \pm 0.03$\,Hz, $0.33 \pm 0.09$ and $0.28 \pm 0.04$ for these three observations. The fit gives a quite satisfactory result: a reduced chi-squared $\chi^2_{\rm red}$ of 0.99/0.97/1.45 for 55/59/55 degrees of freedom. The results show that there are no frequency differences between the three given energy bands for Swift J1658.2--4242 at the frequency of 1.40\,Hz.

\subsection{The features of QPO}
\label{subsect:QPOFeatures}
\begin{figure}[!ht]
  \centering
   \includegraphics[width=0.95\columnwidth]{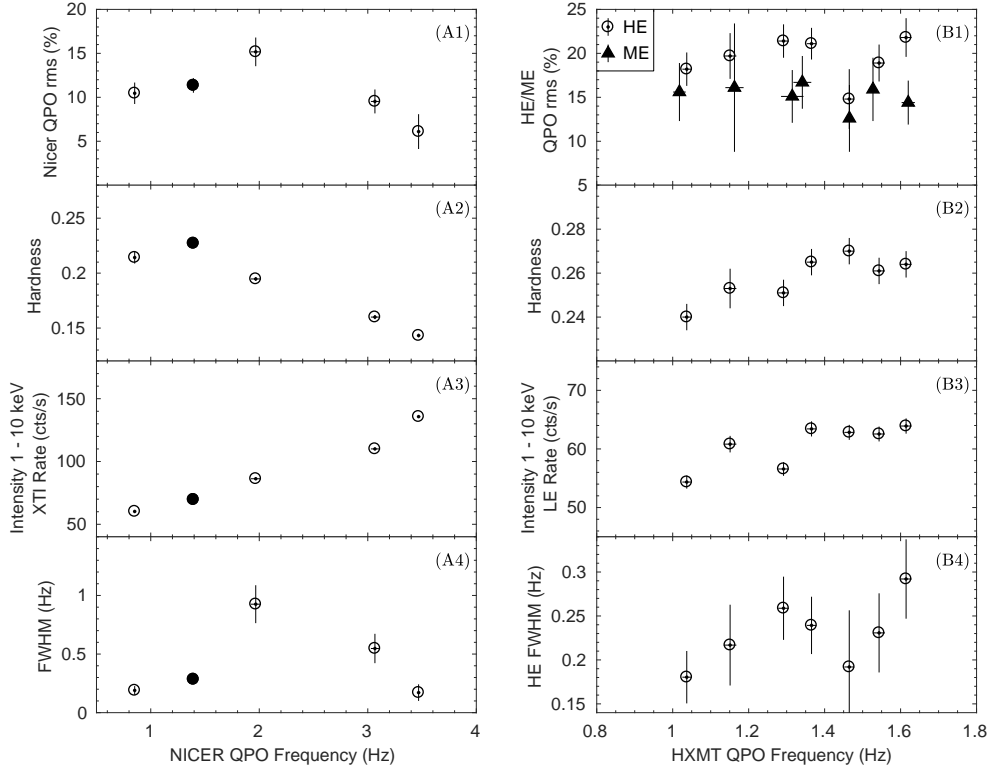}
   \caption{The QPO fractional rms amplitude, hardness, the intensity and FWHM as functions of QPO frequency for Swift J1658.2--4242. The left four panels correspond to {\it NICER} observations and the right four panels correspond to {\it Insight}-HXMT observations. The QPO rms of XTI and ME/HE are given in 1--10\,keV, 10--25\,keV, 25--50\,keV energy bands, respectively. The {\it NICER} hardness is defined as the count rate ratio between 6--10\,keV and 1--6\,keV. The {\it Insight}-HXMT hardness is calculated using the same energy bands with XTI by LE. The filled dots in {\it NICER} panels correspond to observation 1200070104 which was performed on MJD 58169. The QPOs presented in {\it Insight}-HXMT panels are observed on the same day which can be seen as the expansion of {\it NICER} observation 1200070104 in higher energy band.}
   \label{Fig:QPODependenceNicerHXMT}
\end{figure}

Type-C QPOs are often observed in LHS and HIMS, characterized by a strong (up to 20\% rms), narrow ($\nu/\Delta\nu\ge10$) and variable peak whose centroid frequency and intensity vary dramatically in a few days \citep{Motta2015}. Four spectral/timing states (LHS, HIMS, SIMS, HSS) of Swift J1658.2--4242 can be classified from the HID and HRD. The presence of Type-C QPOs observed by {\it NICER} (see Table~\ref{tab:QPOInfoNicer}) characterizes the HIMS of the outburst. The left 4 panels (A1, A2, A3, A4) of Fig.~\ref{Fig:QPODependenceNicerHXMT} present the QPO rms, hardness, intensity and FWHM as function of QPO frequency observed by {\it NICER}. The QPO rms is calculated in the energy band 1--10\,keV. The QPO rms and FWHM increased when the QPO frequency was lower than 2\,Hz and then decreased with the increasing QPO frequency. Similar to those observed black-hole transients \citep{Tomsick2001, Belloni2005, HuangYue2018}, the QPO frequency has positive correlation with the intensity and negative correlation with hardness.

The right 4 panels (B1, B2, B3, B4) of Fig.~\ref{Fig:QPODependenceNicerHXMT} present the QPO rms, hardness, intensity and FWHM as a function of QPO frequency as observed by ME/HE. The QPO rms measured by ME and HE are in the energy bands 10--25\,keV and 25--50\,keV, respectively. As mentioned above, these observations were performed on MJD 58169, the same time as {\it NICER} observation (ID: 1200070104; the filled dot in the left panels of Fig.~\ref{Fig:QPODependenceNicerHXMT}). These observations show the QPO rms, hardness and FWHM increased slightly with increasing QPO frequency except for certain observations (P011465800105, P01146580010506).

\begin{figure}[!ht]
  \centering
   \includegraphics[width=0.95\columnwidth]{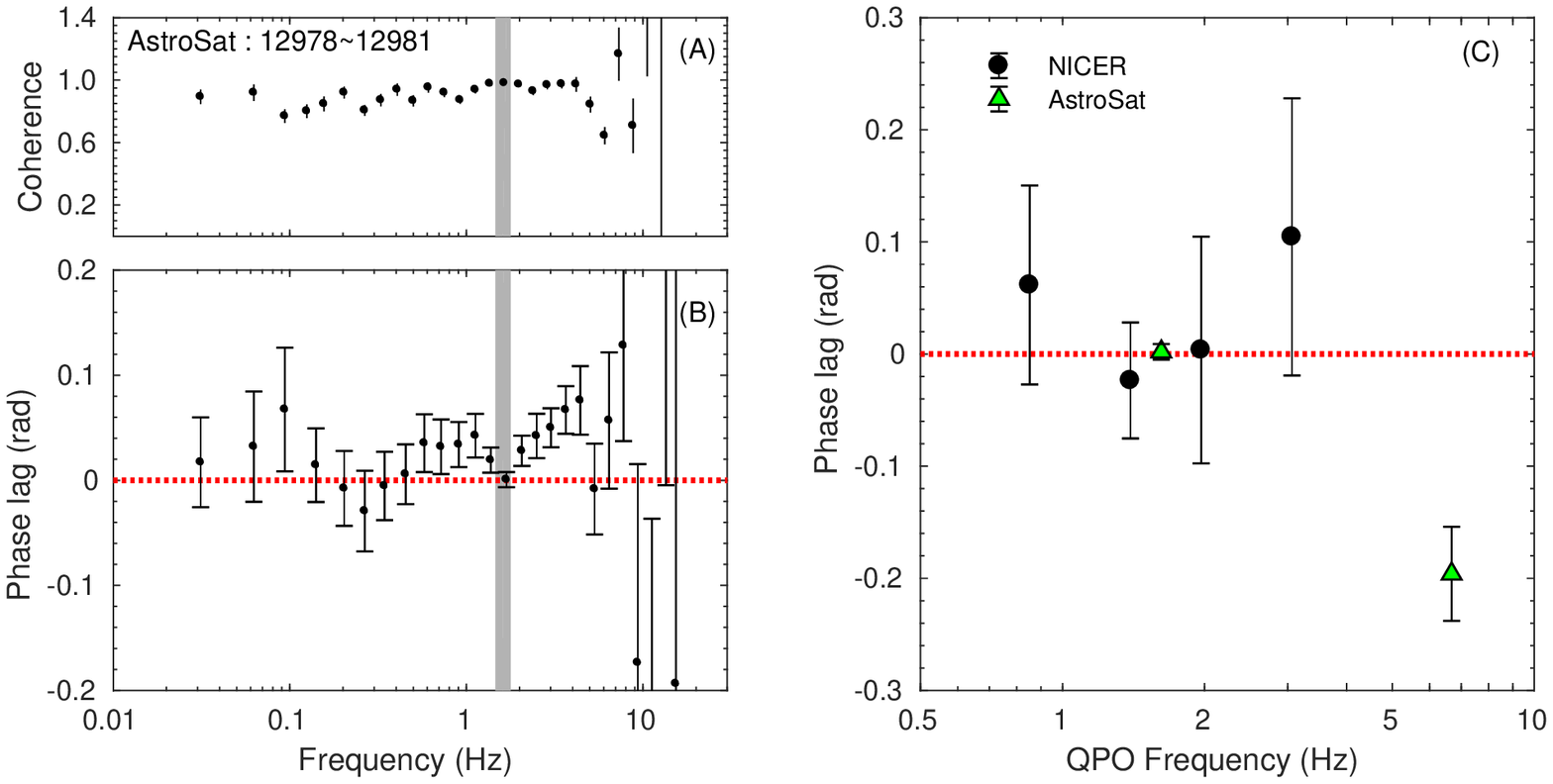}
   \caption{Sample coherence function, phase lag spectrum of {\it AstroSat} observation and phase lag as a function of QPO frequency for Swift J1658.2--4242. Panel (A/B) show the intrinsic coherence function and phase lag in the Fourier domain computed from the {\it AstroSat} merged observations from 12977 to 12981, respectively. The coherence function and phase lag are calculated between the lightcurves corresponding to the 3--6 and 6--10\,keV energy bands. The grey area marks the area of QPO frequency. Panel (C) shows the phase lag (6--10\,keV VS. 3--6\,keV) as a function of QPO centroid frequency using the observations of {\it NICER} and {\it AstroSat}. The four {\it NICER} points in it correspond to observations 1200070103$\sim$1200070106. The observation 1200070107 is not presented here for the low signal-to-noise ratio. The two green triangle points of  AstroSat combine the orbits 12977$\sim$12981 and 13141$\sim$13142, respectively.}
   \label{Fig:SampleLagSpec}
\end{figure}

In order to perform the phase lag analysis, first we computed the intrinsic coherence function of Swift J1658.2--4242 in the QPO-detected observations. As an example, panel (A) of Fig.~\ref{Fig:SampleLagSpec} presents the coherence function between the 6--10 and 3--6\,keV energy band lightcurves of {\it AstroSat} observations combining the orbits of 12977$\sim$12981. The details for computing intrinsic coherence function can be found in the work of \citet{Vaughan1997}. The gray area in this panel marks the range of the QPO. Poisson noise and uncertainties in the {\it AstroSat} and {\it NICER} make the intrinsic coherence estimates above $\sim$10\,Hz unreliable. However, at frequencies below 10\,Hz including the broad band noise ($\sim$0.1--0.6\,Hz) and the QPO frequency, the intrinsic coherence function is essentially unity. This reveals the coherent nature of the two energy bands. The corresponding phase lag spectrum is presented in panel (B) of Fig.~\ref{Fig:SampleLagSpec}. The phase lags at broad band noise including the QPO is close to zero in the observation. While it shows hard lag between 0.8--5\,Hz, shaping a `dip' around the QPO frequency. In  panel (C) of Fig.~\ref{Fig:SampleLagSpec}, we present the QPO phase lag (6--10\,keV VS. 3--6\,keV) as a function of its centroid frequency (0.85--6.68\,Hz) through the whole outburst using the observations of {\it NICER} and {\it AstroSat}. The QPO phase lags maintain zero-some when the centroid frequencies are below 4\,Hz. The observation of {\it AstroSat} shows a soft phase lag at 6.68\,Hz.

\begin{figure}[!ht]
  \centering
   \includegraphics[width=0.95\columnwidth]{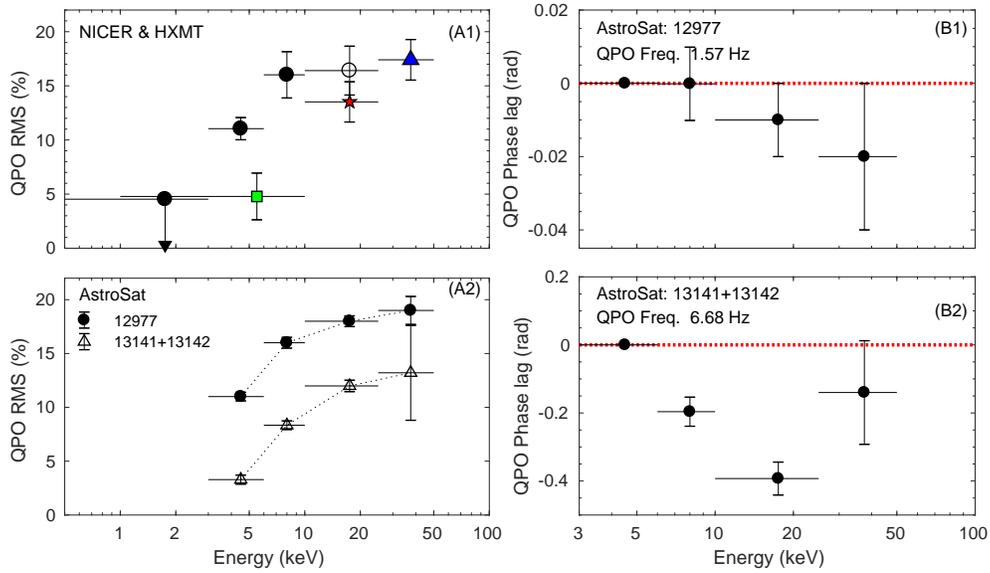}
   \caption{QPO rms and QPO phase lag as function of photon energy in the case of Swift J1658.2--4242. The QPO rms spectrum presented in panel (A1) is calculated from the simultaneous observation by {\it Insight}-HXMT and {\it NICER} presented in subsection~\ref{subsect:SimuObservation}. The black dots, green squares, red star and blue triangle in panel (A1) correspond to the observations of XTI/LE/ME/HE, respectively. The empty circle is the result of ME taking into account of contamination sources (see subsection~\ref{subsect:Contamination}).The energy ranges are 0.5--3\,keV, 3--6\,keV, 6--10\,keV, 10--25\,keV and 25--50\,keV. For comparison, panel (A2) shows the rms spectrum calculated by the nearly simultaneous observation of {\it AstroSat} (black dots). We also checked the rms-energy spectra at the QPO frequency 6.68\,Hz (empty triangles). The right two panels present the QPO phase lag spectra observed by {\it AstroSat}. The QPO frequencies correspond to panel (B1, B2) are 1.57\,Hz and 6.68\,Hz, respectively. In the phase lag spectra, the reference energy band is 3--6\,keV.}
   \label{Fig:SpecRMSSimu}
\end{figure}

Using the simultaneous observations (presented in subsection ~\ref{subsect:SimuObservation}), we present the relation between QPO rms and photon energy of Swift J1658.2--4242 in panels (A1, A2) of Fig.~\ref{Fig:SpecRMSSimu}. Panel (A1) combines the results obtained from the simultaneous data of {\it NICER} and {\it Insight}-HXMT. For comparison, panel (A2) uses the observation performed by {\it AstroSat}/LAXPC (orbit number 12977) nearly at the same time. Also we display the rms-energy spectrum of {\it AstroSat} observation combined from orbit 13141 and 13142 whose QPO frequency increased to 6.68\,Hz in  panel (A2). Considering the low flux of Swift J1658.2--4242, we divided the 0.5--50\,keV into five energy bands to ensure the significance of QPOs. The detailed values are summarized in Table~\ref{tab:RMSInfoAllDet}. As shown in the rms-energy spectra, the QPO rms increases dramatically with increasing photon energy when below 10\,keV. Compared with 0.5--10\,keV, the QPO rms increases slowly in 10--50\,keV energy band. Panel (B1, B2) present the QPO phase lag-energy spectra (6--10\,keV VS. 3--6\,keV) observed by {\it AstroSat}, in which the corresponding QPO frequencies are 1.57\,Hz and 6.68\,Hz. As can be seen in panel (B1), the QPO phase lags in each higher energy band are very close to zero within the $1\,\sigma$ confidence range. While when the QPO frequency reaches 6.68\,Hz, the phase lag becomes softer with increasing photon energy as can be seen from panel (B2). The deviation of the lag in 25--50\,keV from the decreasing trend may result from the low coherence.

To show it more clearly, we list the QPO rms in each energy band observed by the three instruments in Table~\ref{tab:RMSInfoAllDet}. In 3--6\,keV and 6--10\,keV energy bands, the QPO rms measured by LAXPC and XTI correspond to each other very well. In 25--50\,keV, the QPO rms measured by LAXPC and HE is consistent with each other within $1\,\sigma$ confidence range. While in 10--25\,keV, the QPO rms measured by ME is about $25\%$ lower than that of LAXPC, with larger uncertainty and the same problem (even worse) also exists in the results of the LE observations. Tracing this inconsistence, we find a problem of `contamination sources' existing in the FoV of {\it Insight}-HXMT in the case of Swift J1658.2--4242. In short, there exist other sources in the FoV of {\it Insight}-HXMT which should have been a part of the background. This situation could not be modeled in the background software according to its basic principles therefore resulted in the underestimation of `real' background (see subsection~\ref{subsect:Contamination} for more details). This underestimation led to the lower rms when calculating it with equation (\ref{eq:rmsQPO}). After we corrected the background taking into account of contamination sources, the rms measured by ME is consistent with the LAXPC (see the empty circle in panel A1 of Fig.~\ref{Fig:SpecRMSSimu}).

\begin{table}[!ht]
\begin{center}
 \caption{The QPO rms computed from the simultaneous observations of {\it AstroSat}/LAXPC, {\it NICER}/XTI and {\it Insight}-HXMT(ME/HE).}
 \label{tab:RMSInfoAllDet}
 \resizebox{\textwidth}{!}{
 \begin{tabular}{lccccc}
  \hline
  \hline
  Energy Band        &  0.5--3\,keV  &    3--6\,keV      &    6--10\,keV       &     10--25\,keV     &     25--50\,keV         \\
  \hline
  {\it AstroSat} (\%)      &     --       &    11 $\pm$ 0.4  &  16 $\pm$ 0.5      &    18 $\pm$ 0.5    &   19 $\pm$ 1.3         \\
  {\it NICER} (\%)         &   $<$ 4.5    &    11 $\pm$ 1.0  &  16 $\pm$ 2.1      &         --         &        --              \\
  {\it Insight}-HXMT (\%)  &     --       &         --       &       --           &    13.5 $\pm$ 1.9$^{\dagger}$  &   17.4 $\pm$ 1.9       \\
  \hline
  \hline
  \multicolumn{6}{l}{$^{\dagger}$ Rms calculated with the background of contamination sources.}\\
  \multicolumn{6}{l}{$^{\dagger}$ The value is \underline{16.3$\pm$2.3\%} after correction. See details in subsection~\ref{subsect:Contamination}.}\\
 \end{tabular}
 }
 \end{center}
\end{table}

\subsection{Humps Observed by \emph{Insight}-HXMT}
\label{subsect:Humps}
In this part, the humps observed by {\it Insight}-HXMT are presented. In PDSs, humps are the rising signals whose significance cannot reach the level for QPOs. Table~\ref{tab:HumpInfoHXMT} summarizes the humps which are fit with Lorentzians observed by {\it Insight}-HXMT. The rms of humps is calculated with equation (\ref{eq:rmsQPO}). The rms of ME hump is calculated in the energy band 6--30\,keV and HE in 25--50\,keV.

\begin{table}[!ht]
\begin{center}
 \caption{The Hump Information Observed by {\it Insight}-HXMT of Swift J1658.2--4242.}
 \label{tab:HumpInfoHXMT}
  \resizebox{\textwidth}{!}{
 \begin{tabular}{ccccccccc}
  \hline
  \hline
  Obs     & HXMT          &   Central &    ME       &  ME (6--30\,keV)  &        ME        &      ME        & HE (25--50\,keV) &    ME/HE       \\
  Number  & ObsID         &    MJD    & exposure (s)&  Frequency (Hz)   &     FWHM (Hz)    &  Hump rms (\%) & Hump rms (\%)    & Significance   \\
  \hline
  1       & P011465800202 & 58171.27  &    1883     & 2.85 $\pm$ 0.03  & 0.26 $\pm$ 0.10  & 11.0 $\pm$ 3.0 & 13.6 $\pm$ 4.0  & 1.3 / 1.2\,$\sigma$   \\
  2       & P011465800301 & 58172.13  &    2075     & 3.11 $\pm$ 0.06  & 0.62 $\pm$ 0.16  & 12.4 $\pm$ 2.0 & 16.6 $\pm$ 3.0  & 2.3 / 2.0\,$\sigma$   \\
  3       & P011465800302 & 58172.27  &    1949     & 3.02 $\pm$ 0.03  & 0.23 $\pm$ 0.16  &  8.4 $\pm$ 3.2 & 15.0 $\pm$ 3.2  & 1.0 / 1.7\,$\sigma$   \\
  4       & P011465800303 & 58172.40  &    2677     & 3.16 $\pm$ 0.04  & 0.24 $\pm$ 0.08  &  8.1 $\pm$ 1.8 &        -        & 1.6 / -  \,$\sigma$   \\
  \hline
  5       & P011465801902 & 58188.23  &    2382     & 3.83 $\pm$ 0.07  & 0.29 $\pm$ 0.17  &  7.4 $\pm$ 2.8 &        -        & 0.9 / -  \,$\sigma$   \\
  6       & P011465801903 & 58188.39  &    1927     & 4.23 $\pm$ 0.05  & 0.30 $\pm$ 0.15  &  7.8 $\pm$ 2.4 &        -        & 1.2 / -  \,$\sigma$   \\
  \hline
  \hline
 \end{tabular}
 }
 \end{center}
\end{table}

The humps listed here can be separated into two parts according to their appearing date. The first part is the 1--4 observations. This part corresponds to the last two {\it NICER} observations showing QPO features as can be seen in Table~\ref{tab:QPOInfoNicer}. The significance of the QPOs observed by XTI/ME/HE is below 3\,$\sigma$. As mentioned above, we took these two {\it NICER} observations as QPO observations. Hence these humps may be low significance QPOs in higher energy bands. The low significance level may result from the low signal-to-noise level during this time. The second part is 5--6 observations. Humps are found only in ME for these two observations. There is no enough evidence to confirm these two humps as QPOs.

\section{Discussion and Summary}
\label{sect:DiscussionAndSummary}
Using the observations of {\it Insight}-HXMT, {\it NICER} and {\it AstroSat}, we overview the whole outburst of Swift J1658.2--4242 in 2018. Low frequency Type-C QPOs (0.85--6.68\,Hz) are detected. The features of QPOs presented by Swift J1658.2--4242 are similar to many black hole transients. The existing models that attempt to explain the origin of Type-C QPOs are generally based on two different mechanisms: instabilities and geometrical effects \citep{Motta2015}. In the instabilities regimes, a transition layer model is proposed by \citet{Titarchuk2004}. Type-C QPOs are the results of viscous magneto-acoustic oscillations of a spherical bounded transition layer in the above model. Furthermore, \citet{Cabanac2010} proposed a model in which magneto-acoustic waves propagate within a ring-like corona and make it oscillate to explain both Type-C QPOs and the associated noise. In the geometrical effect regimes, the typical physical process is precession. In the Lense-Thirring model, the precessing inner hot accretion flow which is surrounded by a truncated disk accounts for the observed QPOs \citep{Ingram2009}. Recent timing analysis researches indicate that the Lense-Thirring model may explain the origin of Type-C QPOs \citep{Motta2015, Eijnden2017}.

\subsection{QPO rms spectrum}
\label{subsect:RMSSpectrum}
As is shown in Fig. \ref{Fig:SpecRMSSimu}, the QPO rms increases sharply below 10\,keV and mildly in 10--50\,keV energy band. Similar rms-energy spectra for Type-C QPOs were found in MAXI J1535--571~\citep{HuangYue2018}, GRS 1915+10~\citep{Rodriguez2004}, XTE J1859+226~\citep{Casella2004} and XTE J1550--564~\citep{Li2013}. \citet{You2018} computed the fractional rms spectrum of the QPO in the context of the Lense-Thirring precession \citep{Ingram2009}. They found that the rms at higher energy ($E > 10$\,keV) becomes flat when the system is viewed at a large inclination angle.

The fit of {\it NuSTAR} spectrum in the hard state (at the raising stage of the outburst) of Swift J1658.2--4242 shows that the system is viewed at a high inclination angle of $ i=64^{+2}_{-3}$$ ^{\circ}$ and owns a spin parameter of $a^{*}>0.96$ \citep{Xu2018B}. Similar dips are found in the lightcurve of {\it NICER} at the LHS which is close to the end of the outburst. The observed dips in the lightcurves of Swift J1658.2--4242 confirm the high inclination yielded from spectral fitting of {\it NuSTAR}. With these parameters, the Lense-Thirring precession model could explain the observed rms spectrum in Swift J1658.2--4242.

\subsection{QPO phase lags}
\label{subsect:QPOPhaseLag}
As shown in subsection \ref{subsect:QPOFeatures}, a zero constant QPO phase lag within frequency range 0.85--3.5\,Hz and a soft lag at 6.68\,Hz were found. \citet{Eijnden2017} studied the relation between the phase lag at LFQPOs (harmonics and subharmonics as well if exist) and the orbital inclination of black hole transients and found that for Type-C QPOs, both high and low inclination systems possess small hard lag at low QPO frequency ($< 2$\,Hz), while at high frequencies high inclination sources turned to soft lag and low inclination sources become harder. From this point of view, Swift J1658.2--4242 follows the same trend as the high inclination sources (like XTE J1550--564, MAXI J1659--152 \citep{Eijnden2017}), though the QPO whose frequency locates between 4--6\,Hz has not been found with current observations.

In Swift J1658.2--4242, another existing feature is that its QPO lag-energy spectra show different trends depending on the QPO frequency. At low QPO frequencies ($<4$\,Hz), the phase lags stay constant near zero in each energy band. While when the QPO frequency gets above 6\,Hz (6.68 in our case), the phase lags turned softer with increasing energy. The similar feature was found in black hole transients such as XTE J1550-564~\citep{Eijnden2017}. An interesting case is GRS 1915+105 since it shows increasing hard lags at low frequencies~\citep{Qu2010, Reig2000, Eijnden2017}.

\citet{Zhangliang2017} studied the Type-C QPO lag-energy, rms-energy spectra of GX 339--4 and found a critical frequency $\sim1.7$\,Hz. They revealed the important role that the constant reflection played in phase lags when the QPO frequencies were above 1.7\,Hz. Also they suggested that new mechanism might be needed to produce the lags in the observations whose QPO frequencies were below 1.7\,Hz. This remainds us that a critical QPO frequency may exist in Swift J1658.2--4242. While examining the QPO rms-energy spectra at different frequencies (1.57\,Hz and 6.68\,Hz), unlike GX 339--4, we find the shape maintains the same though the variability amplitude in each energy band decreases by $\sim$8\% during the evolution.

Hard lags are much more common than that of zero or soft lags in black hole transients. Three mechanisms were established to explain hard lags: Comptonization models \citep[see e.g.][]{Miyamoto1988, Cui1999, Nowak1999}, reflection models which can produce soft lags as well \citep{Kotov2001} and propagation models \citep[][]{Lyubarskii1997, Kotov2001, Are2006}. Swift J1568.2--4242 does not show the sign of hard lags at all QPO frequencies.

Zero lags at all QPO frequencies are not common in black hole transients compared with hard lags. The high inclination black hole transient GRO J1655-40 shows zero lags at all QPO frequencies ranging from 0.1 to 21\,Hz \citep{Greene2001, Hjellming1995, Eijnden2017}. In the case of GRO J1655--40, a {\it hypersoft} state is reported and interpreted as the sign of a slim disk \citep{Uttley2015}. The jet measurements indicate an extremely high inclination ($85^\circ$) for its inner disk \citep{Hjellming1995}. As a result, the inner accretion flow is possibly obscured by the slim disk. This obscuration could, in the Lense-Thirring precession model, lead to the QPO variations being dominated by quasi-periodic variations of the covering fraction of the inner flow by the slim disk, leading to simple flux variations without corresponding spectral changes that produce the lags \citep{Eijnden2017}. In the case of Swift J1658.2--4242, the spectral fit confirms that the source is highly absorbed. Extra absorption is intrinsic to the source, and most likely originates from obscuring material near the orbital plane of the system \citep{Xu2018B}. From the view of obscuration, Swift J1658.2--4242 may experience the same mechanism as GRO J1655--40 when the QPO frequencies are below 4\,Hz. But it is doubtful that the role of slim disk found in {\it hypersoft} state of GRO J1655--40 can be replaced by the obscuring material near the orbital plane of Swift J1658.2--4242. Another question is that the explanation for soft lag at 6.68\,Hz. Clearly if we adopt the obscuration mechanism, then we will not observe the soft lag except that the geometry of the accretion area changed dramatically. While the similar lag-energy spectrum presented in panel (A2) of Fig.~\ref{Fig:SpecRMSSimu} may indicate very slightly changes in accretion geometry through the evolution of Swift J1658.2--4242 in its HIMS.

In recent years, the detailed lag mechanisms in the framework of Lense-Thirring precession have been developed gradually. \citet{Veledina2013} modeled a precessing inner flow as a precessing Comptonizing ring taking relativistic effects into account~\citep{Eijnden2017}. In their model, both hard  and soft lags at QPO frequency can be produced which is depend on the QPO flux as a function of precession phase. To an observer, the QPO flux varies in a different way in low and high inclination sources. This explains the observed different lag trends in different inclination sources. In the case of Swift J1658.2--4242, the lag-frequency relation and lag-energy spectrum are similar to high inclination sources like XTE J1550--564, H1743--322. Hence we suppose the Lense-Thirring precession model and the mechanism based on it to produce the lags may be the best models to describe Swift J1658.2--4242 by far.

\subsection{QPO rms and contamination sources in {\it Insight}-{\rm HXMT}}
\label{subsect:Contamination}
As has been mentioned in subsection~\ref{subsect:QPOobservations}, the QPO rms calculated by LE/ME is lower than that of XTI/LAXPC in the case of Swift J1658.2--4242. Considering this problem, we examine the background of {\it Insight}-HXMT. It turns out that the contribution of contamination sources in the FoV of {\it Insight}-HXMT could not be modeled in the background software. This underestimation of background resulted in the lower QPO rms calculated from the data of {\it Insight}-HXMT. Hence corrections for the contamination sources are necessary.

\begin{figure}[!ht]
  \centering
   \includegraphics[width=0.95\columnwidth]{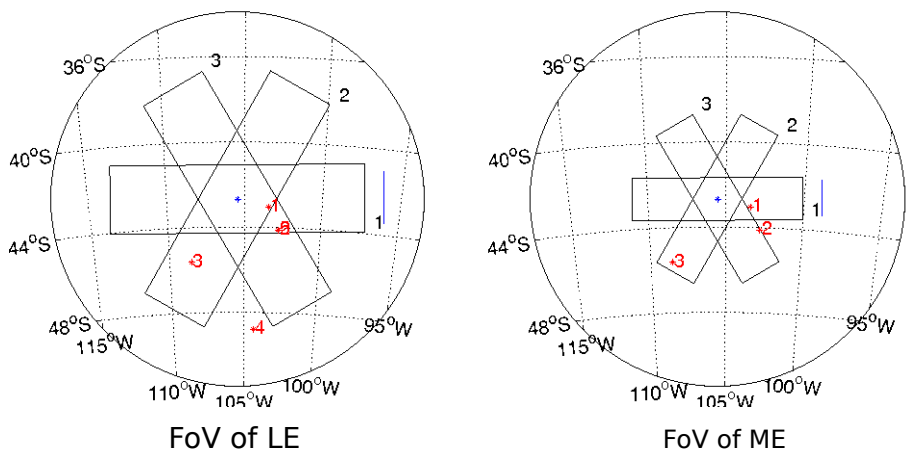}
   \caption{Bright sources in the field of view of LE and ME when observing Swift J1658.2--4242. The red numbers correspond to the positions of contamination sources.}
   \label{Fig:BrightSource}
\end{figure}

Fig.~\ref{Fig:BrightSource} presents the contamination sources in the FoV of LE and ME on MJD 58169 when observing Swift J1658.2--4242, respectively. The supported software \textbf{HXMT Bright Source Warning Tool}\footnote{@\underline{http://proposal.ihep.ac.cn/soft/soft2.jspx}} is available in the official website of {\it Insight}-HXMT. The red numbers correspond to the positions of contamination sources. For LE, the numbers 1--5 correspond to H 1702--429, H 1705--440, GX 340+0, J170248.5--484719, J170855.6--440653, respectively. For ME, the numbers 1--3 correspond to H 1702--429, H 1705--440, GX 340+0, respectively. The three neutron star low massive X-ray binaries (H 1702--429, H 1705--440, GX 340+0) mainly contribute to the flux in lower energy band. Among the contamination sources, H 1720--429 and H1705--440 are classified as atoll sources \citep[see e.g.][]{Mazzola2019, Piraino2007} and GX 340+0 is classified as a Z-source \citep{Seifina2013}. Therefore the data of LE are seriously influenced by the contamination sources. The contamination is too high to correct.

While in the case of ME, the influence of contamination sources is much weaker than that of LE. In the 10--25 keV energy band of ME, the influence of atoll sources H 1705--440 and H 1702--429 on Swift J1658.2--4242 is much weaker than that of GX 340+0. So we focus on the correction of GX 340+0 for ME. Using the fit parameters yielded from two {\it BeppoSAX} observations of GX 340+0 in its normal branch (the most common situation)\citep{Iaria2006}, we simulate the corresponding ME spectrum using \textbf{XSPEC}. The simulated ME spectrum of GX 340+0 gives a count rate of 42.9\,cts/s in 10--25\,keV. Considering the position in the FoV of ME as presented in the right panel of Fig.~\ref{Fig:BrightSource}, we can estimate the observed rate in the following way. First, GX 340+0 can only be observed by the second box of three ME detector boxes. Secondly, the deviation of $\sim2.7^{\circ}$ decreases the observing efficiency to $\sim32\%$ for a typical small FoV ME detector ($1^{\circ}\times4^{\circ}$). Therefor the rate observed by ME can be estimated to lower than $42.9\times\frac{1}{3}\times0.32=4.6$\,cts/s. This can raise the rms of ME in Table~\ref{tab:RMSInfoAllDet} to $16.4 \pm 2.2\%$ (the empty circle in panel A1 of Fig.~\ref{Fig:SpecRMSSimu}) for Swift J1658.2--4242, which is consistent with LAXPC within $1\,\sigma$ confidence range. Note that the rate for contamination sources is not from the simultaneous observation, this background correction may be not suit for the actual situation of GX 340+0 on the observing day.

As for HE, the influence of contamination sources could be ignored in such high energy.

\subsection{Summary}
In this paper, we present the results of a systematical investigation of the timing analysis on the black-hole candidate Swift J1658.2--4242 in its 2018 outburst with the observations of {\it Insight}-HXMT, {\it NICER} and {\it AstroSat}. The main results are summarized as follows:
\begin{itemize}
 \item From the whole outburst, the  HID and HRD are similar to many black-hole transients (e.g. GX 339--4, XTE J1650--500). The four timing/spectral states (LHS, HIMS, SIMS, HSS) could be classified through the HID and HRD.
 \item With the  QPOs detected by three satellites, we present the dependence of QPO rms, hardness, intensity, FWHM on the QPO frequency to investigate the features of HIMS. These correlations are similar to many black-hole candidates like MAXI J1535--571, GX 339--4, Swift J1842.5--1124. The shape of QPO rms spectrum correspond to the simulation computed in the context of Lense-Thirring precession. The QPO phase lag dependence on QPO frequency is similar to high inclination system like XTE J1550--564, which might can be interpreted as the observed flux variation of the Lense-Thirring precessing flow. The QPO rms-energy spectra of Swift J658.2--4242 showing no difference in shape at high/low QPO frequencies may indicate that the accretion geometry experiences no obvious change during HIMS.
 \item Focusing on the simultaneous QPO observations by three satellites in HIMS, QPO frequency does not show differences in three energy bands at $\sim$1.5\,Hz.
 \item As for the humps observed by ME detectors on MJD 58188, we can not tell whether they are QPOs or not with current observations.
\end{itemize}

\section{Acknowledgments}
This work made use of the data from the HXMT mission, a project funded by China National Space Administration (CNSA) and the Chinese Academy of Sciences (CAS). This work made use of the data from the High Energy Astrophysics Science Archive Research Center Online Service, provided by the NASA/Goddard Space Flight Center. This work made use of the data from the {\it AstroSat} mission of the Indian Space Research Organisation (ISRO), archived at the Indian Space Science Data Centre (ISSDC). This work is supported by the National Key R\&D Program of China (2016YFA0400800) and the National Natural Science Foundation of China (NSFC) under grants 11673023, U1838201, U1838115, U1838111, U1838202, 11733009, U1838108.

\section*{References}

\end{document}